%% file: main.tex
\documentclass[prl,aps,twocolumn,preprintnumbers,amssymb,nobibnotes,nofootinbib,10pt]{revtex4-2}
\usepackage{amssymb}
\usepackage{hyperref,graphicx,color,amsmath,dsfont,titlesec,shuffle,tikz}
\usepackage{upgreek}
\usepackage{mathtools}
\input{header_data.tex}

\newcommand{\Tr}[0]{\text{Tr}}
\newcommand{\ah}[0]{\alpha'}
\newcommand{\trp}[0]{\text \Tr \: \phi^3}
\newcommand{\T}[0]{\mathcal{T}^0}
\newcommand{\A}[1]{A^{\ah}\!\!\left(#1\right)}
\newcommand{\M}[1]{M^{\ah}\!\!\left(#1\right)}
\newcommand{\mm}[1]{m^{\ah}\!\!\left(#1\right)}
\allowdisplaybreaks

\begin{document}
	
\title{Quadratic Sum Rules for Scalar Amplitudes}
\author{Karol Kampf}
\author{Ji\v{r}\'i Novotn\'y} 
\author{David Podiv\'in}
\author{Jonah Stalknecht}

\affiliation{Institute for Particle and Nuclear Physics, Charles University, Prague, Czech Republic}

	\begin{abstract}

		\noindent Kleiss--Kuijf and Bern--Carrasco--Johansson relations, and their string monodromy generalizations, have long provided some of the most powerful and useful constraints on partial amplitudes. Motivated by their success, we uncover a new class of quadratic-sum rules that connect scalar amplitudes at different multiplicities. Combining recursion, kinematic shifts, and symbolic regression, we establish them for the nonlinear sigma model and its string-inspired extensions. We then extend these relations to diagonal elements of the inverse Kawai--Lewellen--Tye matrix and subsequently to planar scalar theories with nonderivative interactions and general propagators.
            
	\end{abstract}

	\maketitle
	\vspace{-0.3cm}
    
\section{Introduction}

Recent developments have revealed close connections among colored scalar theories, including the non-linear sigma model (NLSM) and bi-adjoint scalars (BAS). Many essential insights have emerged from combining them into larger unified frameworks. 
Early works on the NLSM derived recursion relations for its amplitudes using the vanishing soft limit \cite{Kampf:2012fn,Kampf:2013vha,Cheung:2014dqa,Cheung:2015ota}. Subsequent work revealed that the soft limits contain ``mixed amplitudes'', where some of the pions are replaced with bi-adjoint scalars \cite{Cachazo2016}. This naturally leads to an extended theory containing both pions and BAS, which provides a natural completion of pion amplitudes to odd multiplicities and admits a set of very efficient recursion relations \cite{Bartsch2024}.

More recent research shows that these scalar amplitudes can be deformed into one another through kinematic shifts. The most well-known example is the \emph{\(\delta\)-shift} \cite{ArkaniHamed2024,ArkaniHamed2024a}, which takes the diagonal BAS amplitudes, known as \(\trp\) amplitudes, and transmutes them to NLSM or mixed amplitudes at even and odd points, respectively. 

\begin{figure}[t]
	\centering
	\includegraphics[width=\linewidth]{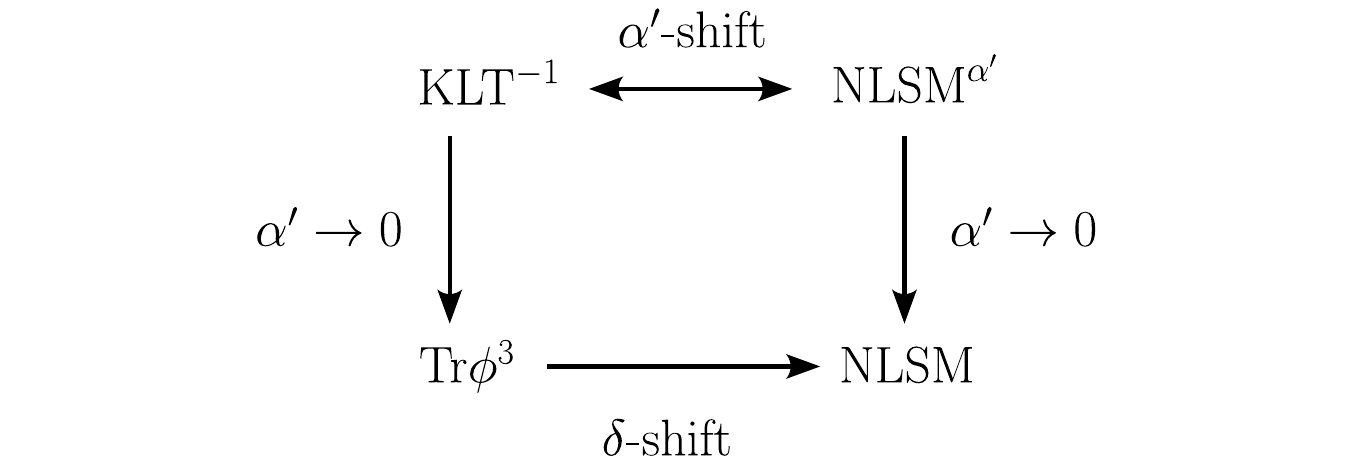}
	\caption{Unity of colored scalar theories.}
	\label{fig:nlsmbasnet}
\end{figure}

While a powerful tool, this process is not invertible, and not all mixed amplitudes can be obtained through this procedure.
These shortcomings can be overcome by considering string-inspired \(\ah\)-completions of both theories. Historically the most studied $\ah$-completion of NLSM is \emph{semi-abelianized Z-theory} \cite{Carrasco2017}. However, more recent work has focused on abelianizing the inverse Kawai--Lewellen--Tye (KLT) kernel \cite{Kawai:1985,Bartsch2025}. The latter construction admits a different kinematical deformation, the \(\ah\)-shift, which provides an invertible transformation between inverse KLT and stringy pions, and hence demonstrates complete equivalence of both theories. These unifications are illustrated in Fig. \ref{fig:nlsmbasnet}. 

In this work, we uncover quadratic-sum rules among the amplitudes within this extended framework. Consistency conditions between Berends--Giele-type recursions \cite{Berends1987,Bartsch2026a} and the \(\delta\)-shift reveal an initial set of sum rules involving pion and mixed amplitudes. 
This prompts us to use symbolic regression to discover complementary pion sum rules, which we further extend to a set of sum rules valid for their \(\ah\)-completions. Using the unity of scalar theories, we derive an equivalent statement for inverse KLT elements. A functional proof of these statements highlights applicability to a broad range of scalar theories without derivative interactions, which can be summarized by the formula
\begin{align}\label{first}
    \sum_{k=2}^{n-1}\Big[\Delta(X_{1k})-\Delta(X_{kn})\Big]A(1\ldots k)A(k\ldots n)=0.
\end{align}
Here \(\Delta\!\left(p^2 \right)\) is a propagator in momentum space, and \(A\) denotes a color--ordered amplitude. We provide several proofs of these sum rules. We also discuss generalizations of these sum rules to theories with spin, derivative interactions, and strings.

Let us compare our sum rules with other relations known in the literature. In the context of scattering amplitudes, the most prominent relations are the linear Kleiss--Kuijf \cite{Kleiss1989} and Bern--Carrasco--Johansson \cite{Bern2008} relations. Their power is illustrated, for example, in NLSM, where they reduce the tree-level amplitude basis \cite{Chen:2013fya} and help constrain both the leading-order amplitudes \cite{Carrasco:2019qwr} and their higher-derivative corrections \cite{Kampf:2021jvf,Brown:2023srz}. Beyond these linear relations, quadratic relations have also been identified, although in more specialized settings. Examples include relations among gauge-theory amplitudes \cite{Bjerrum-Bohr:2010mtb} at fixed multiplicities and, more recently, relations for NLSM currents \cite{Li:2026lqs}.

\section{Review of Colored Scalars}
\paragraph{Notation and Conventions.}
We will be interested in tree--level color-ordered scattering amplitudes in scalar theories. The natural kinematic variables to describe these amplitudes are the \emph{planar Mandelstam variables}
\begin{align}
    X_{ij}=(p_i+p_{i+1}+\ldots+ p_{j-1})^2.    
\end{align}
These $X_{ij}$-variables label the poles of partial amplitudes and form a basis for kinematic space in general dimensions. Since the ordering of momenta is fixed, we can interpret the amplitudes abstractly as functions of particle labels. Concretely, for some ordering $\sigma$ on $n$ particles, we denote
\begin{align}
    m(\sigma)\equiv m(1,2,\ldots,n)|_{X_{ij}\to X_{\sigma(i)\sigma(j)}}.    
\end{align}
The advantage of viewing amplitudes as functions of labels rather than of momenta is that the $X_{ij}$ may be treated as independent formal variables. This allows us to interpret the same expressions as evaluated on higher-point kinematics, and we can use the same notation for off-shell objects such as loop integrands. 

We will make use of this notational freedom in one important way. Some of the formulas below will contain boundary terms in which two-point functions appear together with their propagator. Setting $m(i,i+1)=X_{ii+1}$, both factors will vanish in the on-shell limit, but their ratio approaches unity\footnote{For $\ah$-complete functions, we instead take the two-point functions to be $\mm{i,i+1}=\tan(\pi\ah X_{ii+1})$.}. 
This will help streamline various formulas into a single sum, which would otherwise require isolating the two-point cases into separate terms.

The main objects we will encounter are NLSM, mixed, and $\trp$ amplitudes. For further clarity, we shall denote pure pion amplitudes by \(A\), mixed amplitudes by \(M\), and $\trp$ amplitudes by $m$. Throughout this text, we exclusively consider mixed amplitudes with three adjacent scalars placed on labels \(1,n-1,n\), or, explicitly, \(M\left(\phi \pi \ldots \pi \phi \phi\right)\). 

\paragraph{$\ah$-Completions.}
We will make use of certain stringy $\ah$-extensions of the above field theory amplitudes, which contain these  amplitudes in the $\ah\to0$ limit. The specific class of $\ah$-extensions is motivated by the \emph{inverse KLT kernel}, first studied by Mizera in \cite{Mizera2017a,Mizera2017}, where it was shown that its elements are simple functions of Mandelstam variables. For the purposes of this text, we only consider the diagonal elements, which we denote \(\mm{1\ldots n}\equiv m^{\ah} \!\!\left(\mathds{1}|\mathds{1}\right)\). Explicit examples of these functions are
\begin{align}
	&\mm{123}= 1,\quad \mm{1234} = \frac{1}{t_{13}} + \frac{1}{t_{24}},\\
	&\mm{12345} = 1 +\left(\frac{1}{t_{13} t_{35}} + \text{cyc}.\right),
\end{align}
where we use the notation $t_{ij} \equiv \tan \left(\ah \pi X_{ij}\right)$, and $+\text{cyc}$ denotes the sum over all five cyclic permutations.

It is possible to obtain stringy NLSM amplitudes from the inverse KLT elements via the \(\ah\)-shift \cite{Bartsch2025}
\begin{equation}\label{Ashift}
	\ah \to \ah/2, \qquad  X_{\text{even}} \to X_{\text{even}} \pm {1}/{\ah}.
\end{equation}
Here, \(X_{\text{even}}\) specifies $X_{ij}$ with $|j-i|$ even. 
Examples of these amplitudes are
\begin{align}
	&\A{1234} = -\tau_{13} - \tau_{24},\\*
	&M^{\ah} \!\!\!\left(12345\right) = 1 -\frac{\tau_{13} + \tau_{24}}{\tau_{14}} - \frac{\tau_{24} + \tau_{35}}{\tau_{25}} + \tau_{13}\tau_{35},
\end{align} 
where we have defined \(	\tau_{ij} = \tan \left(\ah \pi \frac{X_{ij}}{2}\right)\).
We stress that the \(\ah\)-shift is reversible, which will be essential later in this text.

To close this section, let us mention that for field theory NLSM amplitudes it is sufficient to start with \(\trp\) amplitudes and employ \(\delta\)-shift defined as \cite{ArkaniHamed2024,ArkaniHamed2024a}
\begin{equation}\label{delta}
	X_{o,o} \to  X_{o,o} - \delta, \qquad X_{e,e} \to X_{e,e} + \delta, 
\end{equation}
where \(X_{o,o},\,X_{e,e}\) are \(X\)-variables with any two odd and any two even labels, respectively. The NLSM amplitudes are then obtained as limits as \(\delta\) goes to infinity:
\begin{align}\label{delta_nlsm}
	&\lim\limits_{\delta \to \infty} \delta^{2n-2}m_{2n}\left(\delta\right) = A_{2n},\\	&\lim\limits_{\delta \to \infty} \delta^{2n-2}m_{2n+1}\left(\delta\right) = M_{2n+1} . 
\end{align}
In other words, by taking \(\delta\) to infinity, we recover pion amplitudes at even-point and mixed amplitudes at odd-point kinematics. Unlike its stringy cousin, the \(\delta\)-shift is generally not invertible.

\section{Mixed Amplitudes Sum Rules}
We begin our exploration by considering the Berends--Giele recursion for even-point \(\trp\) amplitudes \cite{Bartsch2024}, to which we apply the \(\delta\)-shift \eqref{delta}. Expanding around large \(\delta\) yields
\begin{equation}
	m_{2n}\left(\delta\right) = \delta^{3-2n} S_1 + \delta^{2-2n}S_2 + \mathcal{O}\!\left(\delta^{1-2n}\right).
\end{equation}
Comparing against known asymptotic behavior \eqref{delta_nlsm} we see that \(S_2\) is the expected pion amplitude, whereas \(S_1\) has to vanish.
 Writing \(S_1\) explicitly yields the first sum rule
\begin{equation}\label{MSR}
	\sum_{k \; \text{even}}
	\frac{A(1\ldots k) M(k\ldots n)}
	{X_{1,k}} -
	\sum_{k \;\text{odd}}
	\frac{M(1\ldots k) A(k\ldots n)}
	{X_{k,n}}=0.
\end{equation}
A simple example for \(n = 6\) reads
\begin{equation}
	M\!\left(12345\right) - M\!\left(23456\right) = \frac{A\!\left(1234\right)}{X_{14}} - \frac{A\!\left(3456\right)}{X_{36}}.
\end{equation}

The natural next step is to consider \(m_{2n+1}\left(\delta\right)\), however, its \(\delta\) expansion has no super-leading terms and hence does not yield an additional sum rule. Consequently, the presented sum rule only covers the case of even-point kinematics where the highest multiplicity amplitudes present are mixed amplitudes.
\section{Symbolic Regression and Pion Sum Rules}

In the previous section we have described an analytical derivation of a sum rule for mixed amplitudes. Although the same line of reasoning fails to provide a pion sum rule, we may nevertheless look for one. While an exhaustive search over all rational functions is intractable, recent developments in machine learning provide a potential way forward: \emph{symbolic regression}. Given the previous expression, we are motivated to look for \( A\left(1 \ldots n-1\right) - A\left(2 \ldots n \right)\) as a simple function of lower point amplitudes and \(X\)-variables. Here, we proceed by employing symbolic regression and its implementation in \texttt{PySR} \cite{Cranmer2023}. Symbolic regression has seen use in the context of physics as a tool for rediscovering physical laws, a popular example being AI Feynman \cite{Udrescu:2019mnk}. For a recent scattering amplitudes application see \cite{Moynihan2026}, where symbolic regression is used to rediscover the double copy relation of gravity and Yang-Mills.

In our case, we are looking for
\begin{equation}
	A\left(1 \ldots n-1\right) - A\left(2 \ldots n \right) = F\left(A,M,X,\times , + , - ,\div \right).
\end{equation}
That is, we are looking to express the left-hand side as a rational function in lower-point pion and mixed amplitudes and \(X\)-variables, with the mass dimension fixed to match pion amplitudes. 

Indeed, we find the desired sum rule for pion amplitudes
\begin{multline}\label{PSR}
	\sum_{k\;\text{odd}}(X_{k,n} - X_{1,k})M(1 \ldots k) M(k \ldots n)
	\\ +\sum_{k\;\text{even}}(X_{k,n} - X_{1,k}) \frac{A(1\ldots k) A(k \ldots n)}{X_{1,k} X_{k,n}} = 0.
\end{multline}
Symbolic regression, with some manual help, discovered these expressions up to \(n = 9\). 
An explicit example for \(n=7\) reads
\begin{multline}
	A(123456) - A(234567) = \:(X_{57}-X_{15})M(12345) +\\[0.2cm]   \hspace*{0cm} (X_{37} - X_{13})M(34567)  + 
	(X_{47} - X_{14}) \frac{A(1234)A(4567)}{X_{14}X_{47}}.
\end{multline}

\section{KLT Sum Rules and \texorpdfstring{\(\ah\)}{alpha'}-Completion}
So far we have only considered field-theory amplitudes. However, as explained in the introduction, \(\ah\)-completion has proven to be an incredibly useful tool and so we may be inclined towards extending the previous results to stringy amplitudes. Our approach is very similar to that of the previous section: we employ symbolic regression with only two changes. First, the functions under consideration are stringy NLSM and mixed amplitudes, defined by \(\ah\)-shift \eqref{Ashift}. Secondly, we replace \(X\)-variables with both \(t_{ij}\) and \(\tau_{ij}\). Symbolic regression now provides two sets of sum rules
	\begin{multline}\label{FMSR}
	\sum_{k\;\text{even}} \A{1\ldots k} \M{k\ldots n} \left(\frac{1}{\tau_{1k}}+ \tau_{kn} \right)\\-
	\sum_{k\;\text{odd}} \M{1\ldots k} \A{k\ldots n} \left(\tau_{1k} + \frac{1}{\tau_{kn}} \right)=0
\end{multline}
for \(n\) even and
\begin{multline}\label{FPSR}
	\sum_{\text{odd } k}
	\big(\tau_{k,n} - \tau_{1,k}\big)\M{1 \ldots k} \M{k \ldots n}
	\\ +\sum_{\text{even } k}\left(\frac{1}{\tau_{1,k}} - \frac{1}{\tau_{k,n}}\right)
	\A{1\ldots k} \A{k \ldots n}=0
\end{multline} for \(n\) odd.
Again, let us give the basic examples
\begin{align}
	&\A{1234} - \A{2345} = \left(\tau_{35}-\tau_{13}\right) \M{123}\M{345},\notag\\
	&\M{12345}-\M{23456} =\notag\\
    &\hspace{1cm} \frac{\A{1234}}{\tau_{14}} - \frac{\A{3456}}{\tau_{36}}+ \tau_{13}\tau_{35}-\tau_{24}\tau_{46},
\end{align}
for the pion and the mixed amplitudes, respectively. Now we recall the \(\ah\)-shift \eqref{Ashift} and immediately observe that under its reversal, the sum rules \eqref{FMSR},\eqref{FPSR} are mapped to 
\begin{equation}\label{KLTSR}
	\sum_{k=2}^{n-1} \left(\frac{1}{t_{1k}} - \frac{1}{t_{kn}}\right)\mm{1\ldots k} \mm{k\ldots n} = 0.
\end{equation}Thus, we obtain a sum rule for the inverse KLT elements. A simple example for \(n=6\) reads
\begin{multline}
	\mm{12345} - \mm{23456} =\\ \left( \frac{1}{t_{13}}-\frac{1}{t_{36}}\right) \mm{3456}  + \left( \frac{1}{t_{14}}-\frac{1}{t_{46}}\right)\mm{1234}.
\end{multline} 
By virtue of the reversibility of the $\ah$-shift, all these $\ah$-complete sum rules are equivalent to one another. Under consideration of the $\ah\to0$ limit, these stringy relations then imply the field theory sum rules for pions, mixed amplitudes, and $\trp$. What remains is to provide a proof of the KLT sum rule and the rest follows. The relevant proof is provided in a dedicated Appendix.

\section{Extending the sum rules}
The presented sum rules can be extended to a much broader class of theories. 
 Let us consider a theory given by the Lagrangian
\begin{equation}\label{eq:lagrangian}
	\mathcal{L} = -\frac{1}{2} \left< \phi \Delta^{-1} \phi \right> + V\left(\phi\right),
\end{equation} 
where \(\phi=\phi_a T^a\) and \(\left< A\right> \equiv \Tr A\). The only essential properties are that the inverse propagator \(\Delta^{-1}\) vanishes on-shell and that the potential \(V\) does not have any derivatives. The most general case is 
\begin{align}
	\Delta^{-1} = \sum_{k=1}^{\infty}a_k \;\Box^k, \quad V\left(\phi \right) = \sum_{k=3}^\infty b_k \left<\phi^k \right>.
\end{align}
The inverse KLT kernel can be recovered by considering \(\Delta^{-1} = \tan \left(\ah \pi\: \Box\right)\) and \(V\left( \phi \right) = \sum_{k=1}^\infty \frac{C_{k-1}}{2k+1} \left< \phi^{2k+1}\right> \), where \(C_k\) are Catalan numbers. 

Using functional language, the sum rule for \eqref{eq:lagrangian} reads
\begin{multline}\label{funSR}
	\mathcal{M}\left[\phi \right] \equiv \phi \frac{\delta\T}{\delta \phi} -\frac{\delta\T}{\delta \phi}\phi\\+ 
	\left(\Delta \cdot\frac{\delta\T}{\delta \phi}  \right)\frac{\delta\T}{\delta \phi}  - \frac{\delta\T}{\delta \phi} \left(\Delta\cdot \frac{\delta\T}{\delta \phi}  \right)=0,
\end{multline}
where \(\frac{\delta\mathcal{T}^{(0)}}{\delta \phi}\equiv T^a \frac{\delta\mathcal{T}^{(0)}}{\delta \phi^a}\) and \(\mathcal{T}^{(0)}\) stands for tree-level generator of connected amplitudes.
After working through the details, this functional  sum rule produces the familiar expression
\begin{align}
    \sum_{k=2}^{n-1}\Big[{\Delta(X_{1k})}-{\Delta(X_{kn})}\Big]A(1\ldots k)A(k\ldots n)=0.
\end{align}We provide the derivation of the above identities in the Appendix.

As an example, let us consider \( \frac{1}{4} \Tr\:\phi^4\) theory, where we find
\begin{multline}
	A\left(12345 6\right) - A\left( 23456 7\right) =\frac{1}{X_{14}}-\frac{1}{X_{47}}\\
	=\left(\frac{1}{X_{14}}-\frac{1}{X_{47}}\right)A\left(1234\right) A\left(4567\right).
\end{multline}

\paragraph{Derivative Interactions.}
Let us consider a generalization to theories with derivative interactions. A small modification is required, as the sum rule in the functional language reads
\begin{equation}\label{der}
	\mathcal{M}\left[ \phi\right] = \left[J,\frac{\delta S_\text{int}}{\delta \phi}\left[J\right] \right],
\end{equation}
where \(J = \phi + \Delta \cdot \frac{\delta \mathcal{T}^{(0)}}{\delta \phi}\) and \(\mathcal{M}\) is defined by \eqref{funSR}. Equation \eqref{der} can be extended to virtually any colored theory, including Yang--Mills, since the restriction to spin zero only makes the expressions easier to interpret in terms of amplitudes. For example, the sum rule for Yang--Mills theory in Gervais--Neveu gauge reads:
\begin{multline}
	\mathcal{M}\left[A_\mu\right] = \frac{\delta \mathcal{T}^{(0)}}{\delta A_\mu} A_\mu -A_\mu 	\frac{\delta \mathcal{T}^{(0)}}{\delta A_\mu}\\ + \frac{1}{\Box} 	\frac{\delta \mathcal{T}^{(0)}}{\delta A_\mu}  \overset{\longleftrightarrow}{\Box} \frac{1}{\Box}	\frac{\delta \mathcal{T}^{(0)}}{\delta A_\mu}  =-2ig \partial^\nu\left(\left[J_\mu,J_\nu\right] J^\mu\right),
\end{multline} 
where now \(J_\mu=A_\mu - \Box^{-1}\frac{\delta {\cal{T}}^{(0)}}{\delta A^\mu}\).
For scalars we could read these sum rules as statements about amplitudes; here we instead have to understand \(M_\mu=\frac{\delta {\cal{T}}^{(0)}}{\delta A^\mu}\) as a Berends--Giele current with one off-shell leg. 

\section{String theory} 

\begin{figure}
	\centering
	\includegraphics[width=\linewidth]{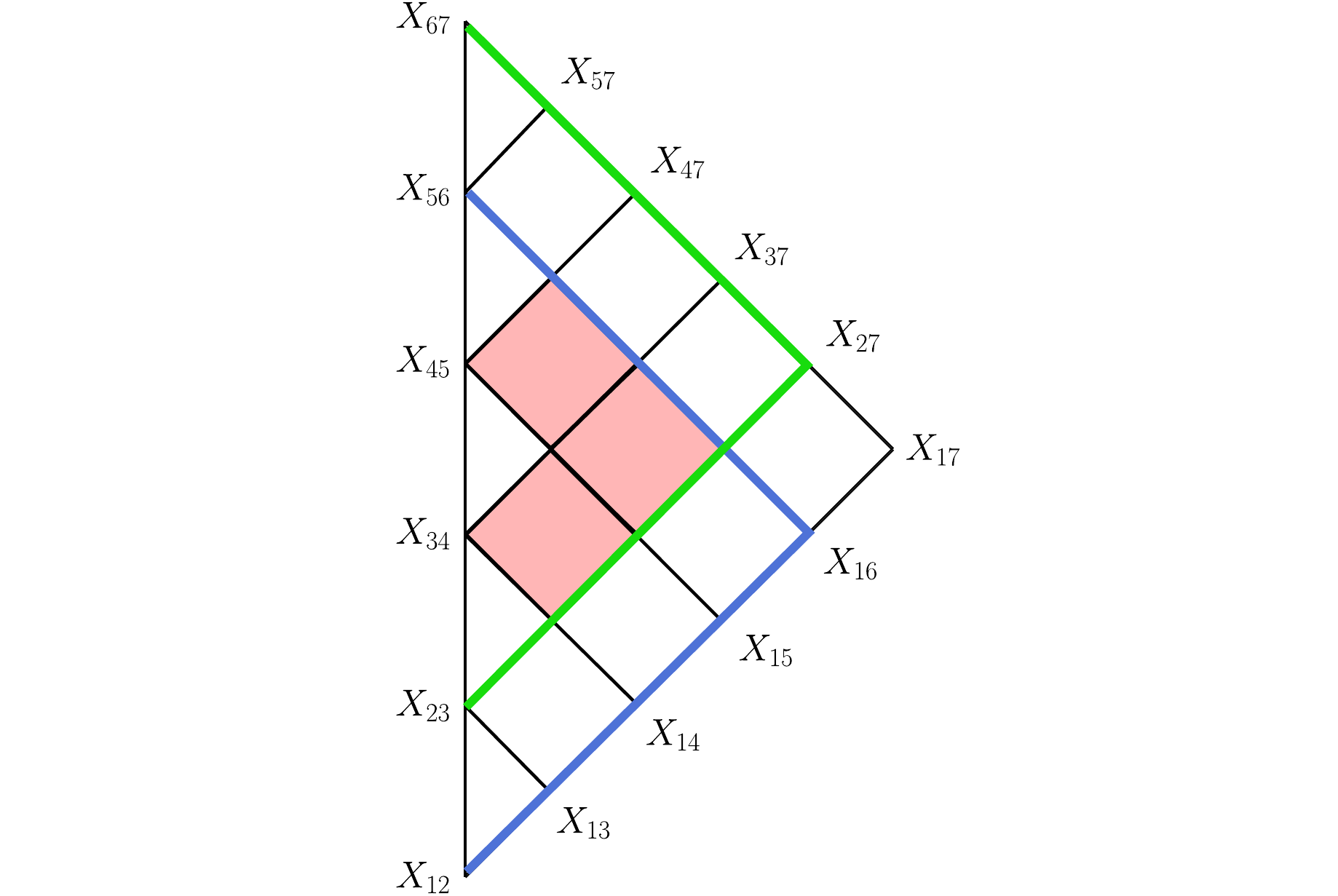}
	\caption{Kinematic mesh for \(n=7\). In minimal kinematics, the red squares are set to zero. The triangles enclosed by blue and green lines symbolize the two amplitudes $\mathcal{A}(123456)$ and $\mathcal{A}(234567)$, respectively.}
	\label{fig:mesh}
\end{figure}

It is natural to ask whether scattering in string theory proper satisfies similar sum rules. We focus on diagonal Z-theory \cite{Broedel2013}
, which is given by the integral over the Koba--Nielsen factor \cite{Koba1969}
\begin{equation}
	\mathcal{A}_n = \int\! \frac{d^{n}\!z}{SL(2,\mathds{R})} \dfrac{\prod_{i<j}\left|z_{ij} \right|^{\ah s_{ij}}}{z_{12}z_{23}\ldots z_{n1}},
\end{equation}
where \(z_{ij} = z_i - z_j\). In order to extend our sum rule to these amplitudes, we first rewrite the $1/X_{1k}-1/X_{kn}$ prefactor as a difference of four-point amplitudes:
\begin{multline}\label{ZSR}
	\mathcal{A}\left(1 \ldots n-1 \right) - \mathcal{A}\left(2 \ldots n \right) = \sum_{k=3}^{n-2}
    \mathcal{A}\left(1\ldots k \right)\mathcal{A}\left(k \ldots n \right)\\  \times
    \left[\mathcal{A}\left(1,k-1, k, k+1 \right) - \mathcal{A}\left(k-1,k,k+1,n \right) \right] .
\end{multline}
In the context of Z-theory, the four-point amplitudes are Euler-Beta functions \(\mathcal{A}_4\left( i,j,k,l\right)= \mathrm{B}\left(\ah X_{ik}, \ah X_{jl}\right)\). The rule (\ref{ZSR}) holds trivially for \(n=5\) and fails for \(n=6\). However, in this case, a closer look reveals that setting \(s_{24} = 0\) recovers the equality. A similar check for \(n=7\) reveals that by setting \(s_{24} = s_{25} = s_{35} =0\) the desired sum rule is again recovered. These kinematics coincide with \emph{minimal kinematics} \cite{Cachazo2021,Early2024,Early2024a,ArkaniHamed2025a,Cao2024a,Cao2024,ArkaniHamed2025b, ArkaniHamed2025}. At this kinematic configuration, the amplitude factorizes completely into Beta functions, or four-point amplitudes. In our case the kinematics is such that all amplitudes in the sum rule factorize in a rather unique way as
\begin{align}\label{Z_factor}
	&\mathcal{A}\left(1\ldots k\right) = \prod_{j=3}^{k-1}\mathcal{A}\left(1, j-1, j, j+1\right), \nonumber\\
	&\mathcal{A}\left(k \ldots n\right) = \prod_{j = k + 1}^{n-2} \mathcal{A}\left( j-1,j,j+1,n \right).
\end{align} 
The specific realization of minimal kinematics we are considering can be described by using \(X\)-variables as
\begin{align}\label{kinZ}
		X_{1j}, \; X_{jn} \text{ are free},\quad X_{ij} = \sum_{k=i}^{j-2}X_{k,k+2} \text{ otherwise}.
\end{align}
These minimal kinematics have a neat visual interpretation in the \emph{kinematic mesh} \cite{Arkani-Hamed:2019vag}, as depicted in Fig. \ref{fig:mesh}. This mesh is a convenient graphical way to organize the planar $X_{ij}$, represented by mesh points, and non-planar $s_{ij}$ variables, represented by diamonds. In this figure, the blue and green triangles can be understood as representing the $n-1$ point amplitudes $m(1,2,\ldots,n-1)$ and $m(2,3,\ldots,n)$, respectively. Our specific minimal kinematics is the requirement that all $s_{ij}$ in the intersection of these two triangles are set to zero.

Interestingly, the sum rule \eqref{ZSR} becomes a rather trivial statement on this minimal kinematics, where the series telescopes term by term. Using factorizations \eqref{Z_factor}, we might also write an equivalent form\footnote{Here the two-point amplitudes are defined to be identically zero.}
\begin{multline}
	\sum_{k=2}^{n-1} \mathcal{A}\left(1\ldots k+1\right) \mathcal{A}\left(k \ldots n\right)\\*-\mathcal{A}\left(1\ldots k\right) \mathcal{A}\left(k-1 \ldots n\right)=0,
\end{multline}
which makes the telescoping transparent. That being said, at no point did we have to specify the four-point Z-theory amplitude \(\mathcal{A}_4\left(1234\right) = \mathrm{B}\left(\ah X_{13},\ah X_{24}\right)\), hence the previous sum rule holds for any fully factorized function \eqref{Z_factor}.
It should be noted that the sum rule for Z-theory works \emph{only} on minimal kinematics, whereas the sum rule holds generally for other theories considered in this work.

\section{Conclusion and Future Directions}

In this letter, we have established a novel set of quadratic sum rules that connect color-ordered amplitudes across different multiplicities and, in some cases, across different particle types. Consistency conditions between recursion relations and kinematic shifts gave rise to the first mixed amplitude sum rules, and symbolic regression provides us with the pion complements. The relations among pion and mixed amplitudes can be derived from the sum rule for the inverse diagonal KLT elements via the unity of $\ah$-complete scalar amplitudes. We provided a proof of these relations via a functional argument, which further establishes that the same sum rules hold for general scalar theories with non-derivative interactions. This provides a common algebraic identity for a broad class of scattering amplitudes, extending beyond the theories accessible through the unity of scalars.

An open question is whether these relations can be interpreted through the lenses of either surfaceology \cite{ArkaniHamed2025c,ArkaniHamed2025a} or positive geometry \cite{ArkaniHamed2018}. Of particular interest are the associahedral grids of \cite{Bartsch2026}, which form a common geometric description for the stringy scalar amplitudes. This could help clarify why pion and mixed amplitudes appear together in the sum rules, and how they relate to the analogous statement for their $\ah$-completions. It is further natural to consider whether similar relations can be found at loop level. Preliminary investigations suggest a possible extension to both field theory integrands and the $\ah$-complete integrands of \cite{Bartsch2026a}. Whether these sum rules hold at all multiplicities and loop orders remains an open problem to which we hope to return in future work.

\medskip
{\it Acknowledgements}: We thank Jin Dong, Shruti Paranjape and Jaroslav Trnka for useful discussions and comments on the draft. This work is supported by GA\v{C}R 27-27733S, GAUK-340126 and OP JAK CZ.02.01.01/00/22\_008/0004632

\bibliographystyle{apsrev4-1}

\bibliography{bib_fix}

\appendix

\onecolumngrid

\section{End Matter}\label{app}

\twocolumngrid

We present two proofs of the sum rules described in this work. The first proof uses functional methods and applies to a very broad range of theories. The second is substantially simpler, more aligned with standard on-shell methods, but also somewhat weaker. 

\paragraph*{Functional proof.} We first present a derivation of the sum rule \eqref{funSR} through a functional Berends--Giele recursion, the details of which can be found in \cite{Bartsch2026a}. Let us consider a theory given by the Lagrangian 
\begin{equation}
    \mathcal{L} = -\frac{1}{2} \left< \phi \Delta^{-1} \phi\right> + V\left(\phi\right) .
\end{equation}
Where \( \left< A  \right> \equiv \Tr A\), the inverse propagator is \( \Delta^{-1} = \sum_{k=1}^{\infty}a_k \Box^k\) and the potential is \(V(\phi)=\sum_{k=3}^{\infty} b_k \left<\phi^k \right>\).  Generator of connected amplitudes \(\mathcal{T}\) is defined by Wick contraction
\begin{equation}
	e^{\frac{i}{\hbar} \mathcal{T}\left[\phi\right]} = \exp \left(
	-\frac{i\hbar}{2} \left< \frac{\delta}{\delta \phi} \cdot  \Delta \cdot \frac{\delta}{\delta \phi}  \right>\right) \exp\left(\frac{i}{\hbar}S_{\text{int}}\left[\phi \right]\right).
\end{equation} Where dots denote integration over spacetime coordinates \(\left< \frac{\delta}{\delta \phi} \cdot  \Delta \cdot \frac{\delta}{\delta \phi} \right> \equiv  \int d^D\!xd^D\!y \frac{\delta}{\delta \phi(x)} \Delta(x,y) \frac{\delta}{\delta \phi(y)}\). Amplitudes are extracted by applying \(\int\! d^D\!x \;e^{i p\cdot x} \frac{\delta}{\delta \phi(x)}\) on \(\mathcal{T}\) for every particle in the desired scattering process and setting \(\phi = 0\). To extract the tree--level generator we expand around \(\hbar \to 0\) 
\begin{equation}
        \mathcal{T} = \T + \mathcal{O}\left(\hbar\right)
\end{equation}
and only consider the leading term \(\T\). The functional form of Berends--Giele recursion takes the form \cite{Bartsch2026a} 
\begin{equation}\label{BGR}
    \frac{\delta \T}{\delta \phi} = \frac{\delta S_{\text{int}}}{\delta \phi} \left[J\right], \quad J = \phi + \Delta \cdot \frac{\delta \T}{\delta \phi}.
\end{equation} Where we have used matrix notation 
\begin{equation}
    \frac{\delta}{\delta \phi} \equiv T^a \frac{\delta}{\delta \phi_a}, \qquad  \phi \equiv \phi^a T^a.
\end{equation} Since \(S_\text{int}\) does not contain derivatives, we observe that \(\frac{\delta S_{\text{int}}}{\delta \phi} = \frac{\partial V(J)}{\partial J}\). 
Putting everything together the sum rule \eqref{funSR} can be written as
\begin{multline}
    \mathcal{M}\left[\phi \right]\equiv \phi \frac{\delta\T}{\delta \phi} -\frac{\delta\T}{\delta \phi}\phi\\+ 
	\left(\Delta \cdot\frac{\delta\T}{\delta \phi}  \right)\frac{\delta\T}{\delta \phi}  - \frac{\delta\T}{\delta \phi} \left(\Delta\cdot \frac{\delta\T}{\delta \phi}  \right)\\=\Bigg[ \phi + \Delta \cdot \frac{\delta \T}{\delta \phi} , \frac{\delta \T}{\delta \phi}\Bigg]= \Bigg[J, \frac{\partial V(J)}{\partial J} \Bigg] = 0.
\end{multline}
Where the first equality is just the definition of a commutator, the second follows from functional Berends--Giele recursion \eqref{BGR} and the commutator vanishes as matrices \(\frac{\partial V(J)}{\partial J} \) and \(J\) commute. 

The functional form of sum rules also allows us to consider theories with derivative interactions. The only new feature is \(\frac{\delta S_{\text{int}}}{\delta \phi} \neq \frac{\partial V(J)}{\partial J}\) and we are thus left with a compact expression
\begin{equation}
    \mathcal{M} \left[ \phi \right] = \Bigg[J , \frac{\delta S_\text{int}}{\delta \phi} \left[ J\right] \Bigg],
\end{equation} where \(J\) is defined in \eqref{BGR}.

Finally, the amplitude version \eqref{first} is obtained by acting \(n-1\)-times with \(\int\! d^D\!x \;e^{i p\cdot x} \frac{\delta}{\delta \phi(x)}\) on \(\mathcal{M}\) and color-stripping.

\paragraph*{Inductive Proof on Amplitudes.}
For a general polynomial interaction $V(\phi)=\sum_{p\ge3}\lambda_p\phi^p$ the Berends--Giele recursion reads
\begin{equation}\label{eq:BG}
  m(1,\dots,n)=\sum_{p=3}^{n}\lambda_p\sum_{\substack{1=j_0<j_1<\dots\\ \dots<j_{p-1}=n}}\prod_{a=1}^{p-1}\frac{m(j_{a-1},\dots,j_a)}{X_{j_{a-1}j_a}}.
\end{equation}
We define
\begin{equation}\label{eq:Bn}
\begin{split}
  B_n\coloneqq{}&m(2,\dots,n)\\
  &+\sum_{k=3}^{n-2}\Big(\frac{1}{X_{1k}}-\frac{1}{X_{kn}}\Big)m(1,\dots,k)\,m(k,\dots,n),
\end{split}
\end{equation}
and show $B_n=m(1,\dots,n-1)$ for all $n\ge4$ by induction on $n$, which implies \eqref{first} for this class of interactions. For $n=4$ the sum is empty and $B_4=m(2,3,4)=\lambda_3=m(1,2,3)$. We now assume $B_p=m(1,\dots,p-1)$ for $4\le p\le n-1$.

\emph{Step 1: $B_n$ does not depend on $X_{in}$, $2\le i\le n-2$.}
By eq.~\eqref{eq:BG}, every term of $B_n$ contains at most one factor $1/X_{in}$ and no other dependence on $X_{in}$. Hence showing that $B_n$ has a vanishing residue as $X_{in}=0$ implies that it is independent of the particle index $n$. The variable $X_{2n}$ does not appear in eq.~\eqref{eq:Bn} at all. For $3\le i\le n-2$, the pole at $X_{in}=0$ arises from factorization of $m(2,\dots,n)$ and of $m(k,\dots,n)$ with $k<i$, and from the explicit $-1/X_{in}$ at $k=i$. This gives
\begin{equation}\label{eq:res}
\begin{split}
  \operatorname*{Res}_{X_{in}=0}B_n
  ={}&m(i,\dots,n)\Big[\,m(2,\dots,i,n)-m(1,\dots,i)\\
  &+\sum_{k=3}^{i-1}\Big(\frac{1}{X_{1k}}-\frac{1}{X_{kn}}\Big)\\
  &\qquad\times m(1,\dots,k)\,m(k,\dots,i,n)\Big]=0,
\end{split}
\end{equation}
where the vanishing follows from the inductive hypothesis.

\emph{Step 2: the limit $X_{in}\to\infty$.}
Since $B_n$ is independent of all $X_{in}$, we may send them to infinity. The terms with $1/X_{kn}$ drop out, and in eq.~\eqref{eq:BG} for $m(k,\dots,n)$ only the terms with $j_{p-2}=n-1$ survive:
\begin{equation}\label{eq:lim}
  m(k,\dots,n)\to\sum_{p}\lambda_p\sum_{\substack{k=j_0<\dots\\ \dots<j_{p-2}=n-1}}\prod_{a=1}^{p-2}\frac{m(j_{a-1},\dots,j_a)}{X_{j_{a-1}j_a}}.
\end{equation}
Inserting eq.~\eqref{eq:lim} into eq.~\eqref{eq:Bn} for $k=2$ and $3\le k\le n-2$ reproduces eq.~\eqref{eq:BG} for $m(1,\dots,n-1)$, with the sum split according to $j_1$: the first term of eq.~\eqref{eq:Bn} is $j_1=2$ (using $m(1,2)/X_{12}=1$), and the term $k$ of the sum is $j_1=k$. Therefore $B_n=m(1,\dots,n-1)$, which completes the induction.

\end{document}

%% file: header_data.tex
\titleformat{\section}{\centering\normalsize\normalfont\bf}{\thesection}{0em}{}
\hypersetup{pdftitle={},pdfcreator={},linkcolor=[rgb]{0.15,0.35,0.75},colorlinks=true,citecolor=[rgb]{0.675,0,0.2},urlcolor=[rgb]{0.15,0.35,0.65}}
\definecolor{nmhv}{rgb}{0.95,0.55,0.55}
\definecolor{mhvblue}{rgb}{0.6,0.6,0.7765}
\definecolor{ampgrey}{rgb}{0.9,0.9,0.9}
\definecolor{hblue}{rgb}{0,0,0.575}
\definecolor{hred}{rgb}{0.575,0.0,0.225}
\definecolor{hgreen}{rgb}{0.0,0.4,0.2}
\definecolor{hteal}{rgb}{0.0,0.545,0.7451}
\definecolor{perm}{rgb}{0.1,0.45,0.85}
\definecolor{unord}{rgb}{0,0,0}
\definecolor{ord}{rgb}{0,0,0.575}
\definecolor{anchorLeg}{rgb}{0.575,0.0,0.225}
\definecolor{fRed}{rgb}{0.48,0.02824,0.18824}